\documentclass [11pt, a4paper, openright, oneside] {article}
\usepackage{float}
\usepackage {amsmath, amssymb, amsfonts}
\usepackage {graphics}
\usepackage {graphicx}
\usepackage {longtable}
\usepackage {pgf,tikz}
\usepackage {float,afterpage} 
\usepackage {verbatim} 
\usetikzlibrary{arrows,automata,shapes,snakes,backgrounds,topaths,petri}
\usepackage {subfigure}
\usepackage{chemarr} 
\usepackage{cite}
\usepackage{lscape}
\usepackage{setspace}
\usepackage{bbm,dsfont}
\DeclareMathOperator{\argmax}{argmax}
\DeclareMathOperator{\argmin}{argmin}
\usepackage{algorithm}
\usepackage{algorithmicx,algpseudocode}

\title{Optimizing Threshold--Schedules for Approximate Bayesian Computation Sequential Monte Carlo Samplers: Applications to Molecular Systems}
\author{Daniel Silk$^\ast$, Sarah Filippi$^\ast$, Michael P.H. Stumpf$^\dagger$\\[6mm]
Centre for Integrative Systems Biology at Imperial College London\\\\
$^\ast$ these authors contributed equally\\
$^\dagger$ to whom correspondence should be addressed, m.stumpf@imperial.ac.uk}
\begin{document}
\maketitle

\abstract{The likelihood--free sequential Approximate Bayesian Computation (ABC) algorithms, are increasingly popular inference tools for complex biological models. Such algorithms proceed by constructing a succession of probability distributions over the parameter space conditional upon the simulated data lying in an $\epsilon$--ball around the observed data, for decreasing values of the threshold $\epsilon$. While in theory, the distributions (starting from a suitably defined prior) will converge towards the unknown posterior as $\epsilon$ tends to zero, the exact sequence of thresholds can impact upon the computational efficiency and success of a particular application. In particular, we show here that the current preferred method of choosing thresholds as a pre-determined quantile of the distances between simulated and observed data from the previous population, can lead to the inferred posterior distribution being very different to the true posterior. Threshold selection thus remains an important challenge. Here we propose an automated and adaptive method that allows us to balance the need to minimise the threshold with computational efficiency. Moreover, our method which centres around predicting the threshold -- acceptance rate curve using the unscented transform, enables us to avoid local minima - a problem that has plagued previous threshold schemes.}

\section{Introduction}
Mathematical models have become powerful tools for both summarising our current biological understanding, and generating novel hypotheses. However, as our models become more ambitious in size and complexity, the computational challenges of a number of tasks such as parameter inference and model validation are increasingly vast. For large, complex or stochastic models, exploring the likelihood surface can be too complicated or numerically too demanding, even though it is possible to simulate the model. For this reason, likelihood-free methods such as Approximate Bayesian Computation (ABC), and its more efficient sequential versions, are becoming increasingly important.
\par
Sequential ABC algorithms proceed by constructing a succession of probability distributions over the parameter space conditional on the simulated data lying in an $\epsilon$--ball around the observed data and use decreasing values of the threshold $\epsilon$ to incrementally approximate the true posterior distribution. While in theory, the distributions (starting from a suitably defined prior) will converge towards the unknown posterior as $\epsilon$ tends to zero, in practice the exact sequence of thresholds can have a great impact on the computational efficiency and success of a particular application. Currently, thresholds are typically chosen as a pre-determined quantile of the distances between simulated and observed data from the previous population, or simply by intuition -- the drawbacks of which are made clear in the results below.
\par
Here we present an automated and adaptive method for threshold choice that is based upon predictions of the future distributions using the unscented transform (UT). Generally known for its use in extending the Kalman filter to non-linear problems, the UT allows the statistics of a Gaussian random variable that has undergone a non-linear transform to be estimated. In combination with Gaussian mixtures, the UT can be used to predict the ABC acceptance rate for any threshold value, and subsequently choose the acceptance thresholds that optimally balances the need to minimise $\epsilon$ with computational efficiency. Further, knowledge of the complete threshold -- acceptance rate curve can enable us to avoid local optima --- a problem that, as is often but perhaps not always acknowledged, has plagued previous threshold schemes. We will show below that this problem is particularly pertinent for schemes that choose threshold schedules from quantiles of the previous population of accepted particles.
\par
The remainder of the paper is organised as follows: we first introduce ABC SMC and use a toy model to discuss the challenges of threshold selection, and in particular, the difficulty of avoiding local optima. We then describe the proposed threshold selection scheme for inference on deterministic models (with Gaussian measurement error); although within an ABC filtering framework, a natural extension allows applications also to stochastic state-space models. Finally we compare the performance of the new adaptive method with various fixed quantile schedules for inference on both toy and biological systems, including two biochemical oscillators.

\section{Adaptive Sequential Monte Carlo methods in Approximative Bayesian Computation}

The aim of ABC is to obtain a good and computationally affordable approximation to the posterior distribution
$$
p(\theta|x^*) \propto  f(x^*|\theta)\pi(\theta),
$$
where $\pi(\theta)$ denotes the prior distribution over the parameter space and $f(x^*|\theta)$ is the likelihood of the observed data $x^*$ for a given parameter, $\theta$.
Rather than evaluating the likelihood directly, which for many real-world problems can be intractable, ABC-based approaches use systematic comparisons between real and simulated data. The main principle consists of comparing the simulated data, $x$, with the real data, $x^*$, and accepting simulations if a suitable distance measure between them, $\Delta(x,x^*)$, is less than a specified threshold, $\epsilon$.
The ABC algorithm thus provides a sample from the approximate posterior of the form,
$$
p(\theta|x^*)\approx p_\epsilon(\theta|x^*)\propto\int f(x|\theta)\; \mathds{1}\left(\Delta(x,x^*)\le \epsilon\right) \pi(\theta) dx\;.
$$
\par
The simple ABC scheme outlined above suffers from the same shortcomings as other rejection samplers: most of the samples are drawn from regions of parameter space which cannot give rise to simulation outputs that resemble the data. Over the past few years many improvements to these algorithms have been proposed that makes ABC inference more efficient: regression-adjusted ABC \cite{Tallmon:2004jp, Fagundes:2007fr, Blum:2010hv}, Markov chain Monte Carlo ABC schemes \cite{Marjoram:2003fn, Ratmann:2007hh}, and ABC implementing variants of sequential importance sampling (SIS) or sequential Monte Carlo (SMC) \cite{Sisson:2007aa, Toni:2009p9197, Beaumont:2009be, DelMoral:2008tm, Drovandi:2011hm}. We will focus here on the last form, which appears to emerge as the most popular framework for complex inferential problems \cite{Liepe:2010p32436}.
\par
\afterpage{
\newpage
\begin{figure}[H]
\center
\includegraphics[width=1.\textwidth]{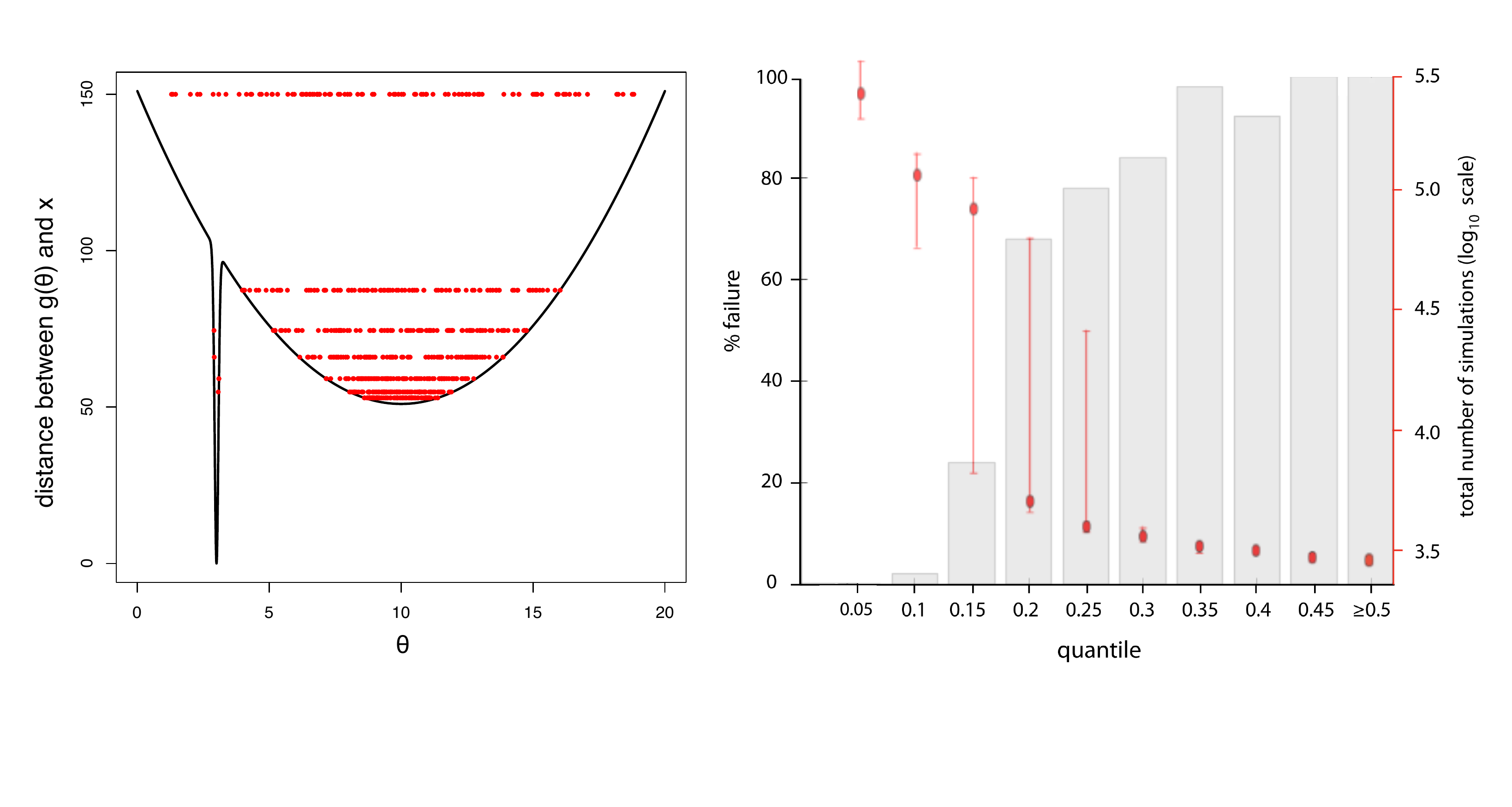}
\caption{\footnotesize{Global and local minima. (a) Plot of the distance between simulated and true data for different parameter values. The model was designed to produce a severe global optimum at the true parameter value $\theta = 3$, and a broad local optimum associated with distances no smaller than 50. The red dots arranged in horizontal lines are the members of successive ABC accepted populations. After the sixth population, parameters near the global optimum are no longer sampled. (b) Plots of the failure rates (grey), and total number of simulations (red) for a range of fixed quantile schedules. The red dots indicate the means over 100 ABC SMC runs, while the upper and lower whiskers are the maximum and minimum values respectively.}}
\label{fig:stuck}
\end{figure}  
}

ABC methods based on SIS or SMC samplers aim to sample from a sequence of distributions, which increasingly resemble the target posterior; they are constructed by estimating intermediate distributions $p_{\epsilon_t}(\theta|x)$ for a decreasing sequence of $\{\epsilon_t\}_{1\leq t\leq T}$. In this article, we focus on the implementation of Toni et al \cite{Toni:2009p9197} and Beaumont et al \cite{Beaumont:2009be} described in Algorithm~\ref{algo:ABCSMC}. This implementation that we will call in the following ABC SMC differs from the ABC SMC algorithm of Del Moral et al \cite{DelMoral:2008tm} and Drovandi et al \cite{Drovandi:2011hm} in a number of points, and proceeds as follows: the first population of \textit{particles} is constructed using the rejection ABC algorithm described above with a sufficiently large value of $\epsilon_1$ such that many particles are accepted: the parameters $\theta$ are drawn from the prior distribution $\pi(\theta)$, and are accepted only if the distance between the simulated and observed data is smaller than $\epsilon_1$. We denote by $\{\theta^{(i,t)}\}_{1\leq i\leq N}$ the set of accepted particles at step $t$, and by $\{x^{(i,t)}\}_{1\leq i\leq N}$ the corresponding simulated data. Each particle $\theta^{(i,t)}$ has an associated weight $\omega^{(i,t)}$; in the first population all weights are equal to $1/N$. For each intermediate population $t$, a parameter $\{\theta^{(i,t-1)}\}_{1\leq i\leq N}$ is sampled from the previous population, $t-1$,  with probability defined by the weights, $\{\omega^{(i,t-1)}\}_{1\leq i\leq N}$, and perturbed using a \textit{perturbation kernel}, $\tilde{\theta}\sim K_t(\cdot|\theta)$; the parameter $\theta$ is then accepted if and only if the distance between the simulated and the observed data is smaller than $\epsilon_t$. These sample, perturbation, simulation and acceptation/rejection steps are repeated until $N$ particles have been accepted. The weight of each particle is then computed as
$$
\omega^{(i,t)} =\frac{\pi(\theta^{(i,t)})}{\sum_{j=1}^N \omega^{(j,t-1)}K_t(\theta^{(i,t)}|\theta^{(j,t-1)})}\; .
$$

The efficiency of the sequential ABC algorithm described above strongly relies on the choice of the perturbation kernel $\{K_t(\cdot|\cdot)\}_t$ as well as the sequence of thresholds $\{\epsilon_t\}_t$. Over the past years adaptive methods to choose perturbation kernels have gained popularity. Beaumont et al \cite{Beaumont:2009be} first suggested to use a componentwise-normal perturbation kernel with an adaptive choice for the variances. Filippi et al \cite{filippi2011optimality} then generalized this approach to a multivariate normal perturbation kernel and compared the efficiency of the ABC SMC algorithm for a selection of adaptive covariance matrices.
\par
Until recently, there has been no systematic way of determining the threshold sequence. The ideal threshold scheme is the one that minimizes the total number of simulations since this is typically the most computationally expensive part of any ABC algorithm. This requires a careful balance between a small number of populations i.e. a rapidly decreasing sequence of thresholds, and a high acceptance rate per round which generally happens if the difference between two consecutive thresholds is small enough. In the following, we denote by $\aleph_t$ the acceptance rate for the round $t$ which is equal to ratio of the population size, $N$, and the number of times the model has been simulated during the $t$-th round. 
Perhaps the most commonly used adaptive scheme for threshold choice is based on the quantile of the empirical distribution of the distances between the simulated data from the previous population, and the observed data (see \cite{Beaumont:2009be,lenormand2011adaptive} and in a different way \cite{DelMoral:2008tm,Drovandi:2011hm}).  The method determines $\epsilon_t$ at the beginning of the $t$-th round by sorting the distances $\{\Delta(x^{(i,t-1)},x^*)\}_{1\leq i\leq N}$ and setting $\epsilon_{t}$ such that $\alpha$ percent of the simulated data $\{x^{(i,t-1)}\}_{1\leq i\leq N}$ are below it, for some predetermined $\alpha$.
\par
A severe drawback of this quantile approach for threshold selection is that the final ABC posterior distribution $p_{\epsilon_T}(\theta|x^*)$ may end up being very different to the true posterior $p(\theta|x^*)$. In particular, if particles are sampled from a large region of parameter space that offers negligible or little support for the posterior distribution, there is a risk of getting stuck in this parameter region if the threshold is selected using a quantile method. As an illustration we consider a toy model where for each $\theta$, the simulated data is  $x=g(\theta)=(\theta-10)^2-100\exp(-100(\theta-3)^2)$. Moreover, we suppose that the true data are generated using the parameter $\theta^*=3$. The support of the posterior distribution should then contain this parameter value. 
\par
Figure \ref{fig:stuck} represents the $L1$ distance between the simulated data $g(\theta)$ and $g(3)$ as a function of $\theta$. In this example, for all the parameter space except in the interval $(2.92, 3.08)$ the distances are larger than $50$. ABC SMC is used on this example, selecting the threshold sequences as the $0.8$ quantile of the previous population's distances. The prior distribution is Gaussian with mean $10$  and variance $10$.  Particles from successive populations are represented by the red dots in Figure \ref{fig:stuck}, with population $t$ aligned along the horizontal line $y=d_t$ where $d_t$ is the maximum distance between $g(\theta^{(i,t)})$ and $g(3)$ for all $1\leq i \leq N$. For example,  $\max_i|g(\theta^{(i,1)})-g(3)|$ being equal to $150$, the first population is represented by dots on the line corresponding to $y=150$. We note that for all $t$ the distributions $p_{\epsilon_t}(\theta|x^*)$ are centred around $10$ and that the true parameter $3$ has a very low probability under the final distribution, where the value of the threshold has converged. Repeating this inference for different values of $\alpha$, leads to very different results (figure \ref{fig:stuck}). The optimal (or at least safe) choice of $\alpha$ thus depends on the data, the model and the prior range. We feel that this problems highlights the potential issues arising in real-world applications. 
\par
The fundamental idea underlying our approach is to predict reliably and cheaply how the acceptance rate depends on the value of the threshold $\epsilon$. For this example, the key is to avoid areas where the acceptance rate is excessively high. Too high an acceptance rate means to overly reward particles that are similar to the ones from the previous population; this in turn will lead to particle populations tending to drift in parameter space, whence broad but shallow local optima are explored more frequently than they ought to be compared to smaller regions that have higher posterior probability.  
\par
Below we present an automated and adaptive method for threshold choice based upon predictions of future distributions. We show that by predicting acceptance rates over all threshold values, we are able to detect possible local minima and choose thresholds so that they are avoided. Furthermore, global knowledge of these acceptance rates allows us to balance threshold reduction and computational cost at each round of the ABC SMC algorithm. We illustrate the method with applications to toy and biological dynamical systems, comparing the performance of our method with various quantile selection schemes.
\afterpage{
\begin{algorithm}[H]
 \begin{algorithmic}[1]
   \caption{ABC SMC algorithm}
   \label{algo:ABCSMC}
   \State {\bfseries input:} a decreasing sequence of thresholds, $(\epsilon_t)_{1\leq t\leq T}$ such that $\epsilon_T=\epsilon$, a data $x$, a sequence of $(K_t(\cdot|\cdot))_{1\leq t\leq T}$
   \State {\bfseries output:} a weighted sample of particles from $p_{\epsilon_T}(\theta|x)$
   \For{ all $1\leq t\leq T$} 
   \State determine the perturbation kernel $K_t(\cdot|\cdot)$ and the next threshold $\epsilon_t$
   \State $i \leftarrow 1$
   \Repeat
   \If{t=1}
   \State sample $\tilde\theta$ from $\pi(\theta)$
   \Else
   \State sample $\theta$ from the previous population $\{\theta^{(i,t-1)}\}_{1\leq i\leq N}$ with weights $\{\omega^{(i,t-1)}\}_{1\leq i\leq N}$
   \State sample $\tilde\theta$ from $K_t(\cdot|\theta)$ and such that $\pi(\tilde\theta)>0$
   \EndIf
   \State sample $y$ from $f(\cdot|\tilde\theta)$
   \If {$\Delta(y,x)\le\epsilon_t$}
   \State $\theta^{(i,t)}\leftarrow\tilde\theta$
   \State $i\leftarrow i+1$
   \EndIf
   \Until{ $i=N+1$}
   \State calculate the weights: for all $1\leq i\leq N$
   \If{$t\neq 1$}
   $$\omega^{(i,t)} \leftarrow\frac{\pi(\theta^{(i,t)})}{\sum_{j=1}^n \omega^{(j,t-1)}K_t(\theta^{(i,t)}|\theta^{(j,t-1)})}$$
   \Else $\quad\omega^{(i,1)}\leftarrow1$
   \EndIf
   \State normalize the weights
	\EndFor
 \end{algorithmic}
\end{algorithm}
}


\section{The threshold -- acceptance rate curve}
The proposed method centres around understanding the acceptance rate $\aleph_t(\epsilon)$ as a function of the threshold $\epsilon$ for the next round of ABC simulations. We ask, if the threshold -- acceptance rate curve were known, how should $\epsilon$ be chosen in order to optimally balance computational efficiency with the need to minimise $\epsilon$, and to avoid getting stuck in regions of parameter space that share little support with the posterior distribution? 
\par
Figure \ref{fig:epsiloncurves} shows the threshold -- acceptance rate curves for a variety of models. 
Although the structure of the proposal distribution and likelihood surface could give rise to anything monotonic increasing, we tend to encounter three main types or curve (shown in figure \ref{fig:epsiloncurves}); concave, convex and sigmoidal. Further, for their interpretation it helps to consider each curve as a combination of convex and concave parts.  A concave shape occurs over a particular range of threshold values, when the majority of particles drawn from the perturbed distribution (with distances relevant for this range) give rise to simulated data that is relatively close to the observed data. The opposite is true for convex shapes. 
\par
\afterpage{
\newpage
\begin{figure}[H]
\center
\includegraphics[width=0.8\textwidth]{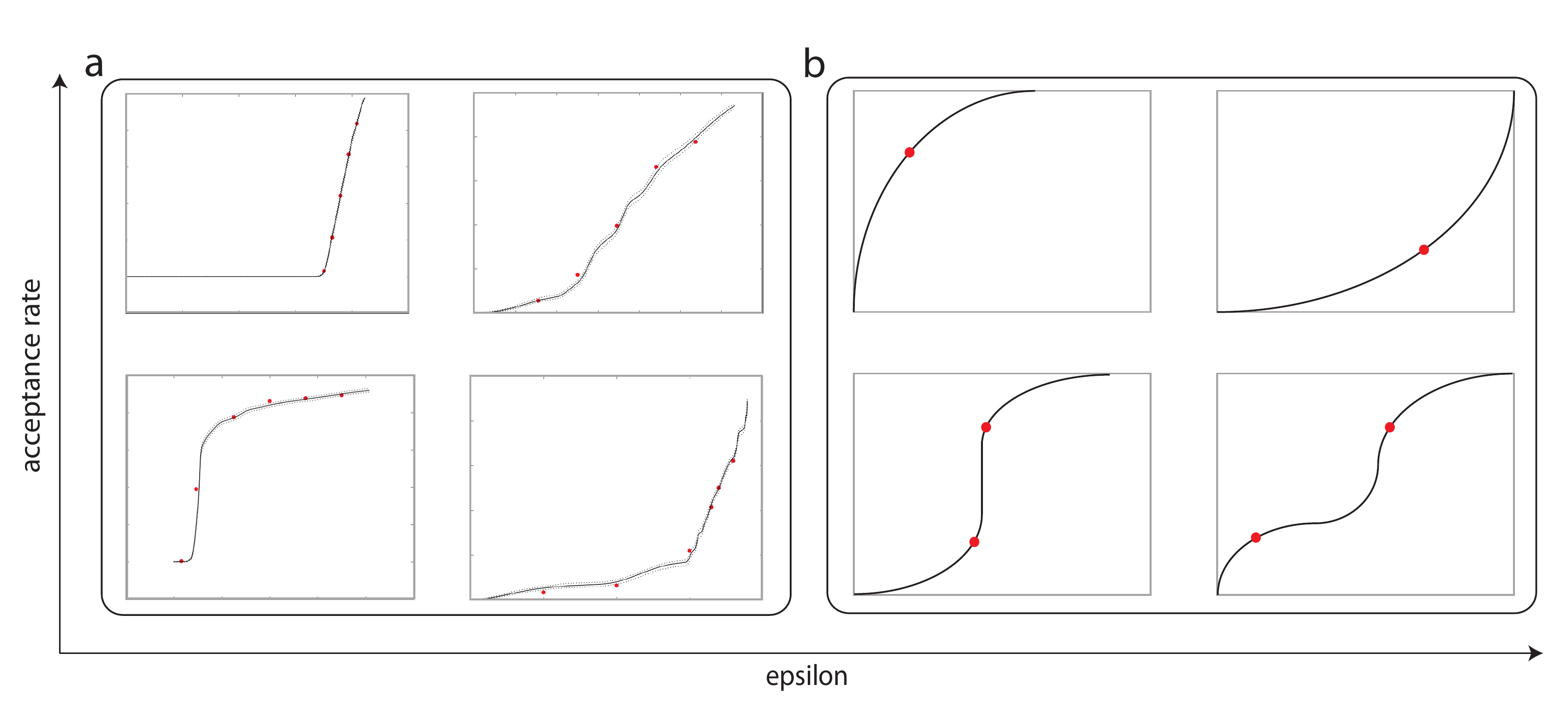}
\caption{\footnotesize{(a)Epsilon--acceptance rate curves for different models (clockwise from top left); a Gaussian p.d.f, a quadratic function, the Repressilator, and the smallest biochemical system exhibiting a Hopf bifurcation. Solid lines indicate mean values of 100 repeat predictions, with dotted lines the minimum and maximum predicted values. Red dots indicate the actual acceptance rate of an ABC run with corresponding threshold values. (b) Typical threshold -- acceptance rate curve shapes. Red dots indicate the threshold values considered by our method - either at the extreme of a convex section to avoid a possible local optimum, or the point $\argmin_\epsilon\Delta((\frac{\epsilon}{\epsilon_{t-1}}, \frac{\aleph_t(\epsilon)}{\aleph_t(\epsilon_{t-1})}),(0,1))$ that balances reduction of $\epsilon$ with computational expense. In the case where multiple threshold values satisfy the latter condition, we select the smallest one.}}
\label{fig:epsiloncurves}
\end{figure}
}
One way the latter can happen is when the perturbed distribution spans a region of parameter space that includes both a sharp global maximum and a broader local maximum of the likelihood. To see this, one can imagine an $\epsilon$--ball expanding about the true data; at first the ball only encompasses a small number of particles that were drawn from very close to the global maximum, corresponding to the low gradient at the foot of the shape. Once $\epsilon$ is large enough we are able to accept the relatively large number of particles sitting in the local maximum, which causes the increase in gradient. This is exactly the scenario described in the toy example above where the ABC SMC algorithm is seen to fail when using various quantile strategies to select the threshold values (figure \ref{fig:stuck}), and is likely to be a common occurrence in biological systems where likelihood surfaces are known to be highly complex \cite{Gutenkunst:2007ww,Erguler:2011bu}. Indeed below we present such an example involving the smallest possible biochemical system that can exhibit an oscillation inducing Hopf bifurcation \cite{Wilhelm:1995uc,Kirk:2008ws}. 
\par
This interpretation of convex shapes as a symptom of sampling from local optima in the posterior suggests  the following criterion for threshold selection: the ABC SMC algorithm can get stuck when a threshold schedule allows particles from a relatively broad local optimum to be accepted too frequently and for successive populations --- at each round of the algorithm we risk not sampling from the correct posterior (i.e. that part of the posterior that account for ``almost all" of the probability mass of the posterior) at all; diffusion of the parameter particles to the incorrect area in parameter space is thus entropically driven. This can be avoided by choosing a threshold that rejects these particles with high probability.  From the threshold -- acceptance rate curve we can identify such a threshold value as one that lies at the bottom of the steep incline, i.e. $\argmax_\epsilon \frac{\partial^2 \aleph_t(\epsilon)}{\partial \epsilon^2}$. For concave shapes the same danger is not apparent from the curve, and we can instead try to balance computational expense against the desired reduction in $\epsilon$. Here we treat the threshold -- acceptance rate curve in the same spirit as we would treat a ROC curve, identifying the optimal ``cut-point'' as that which minimises the distance between $(\frac{\epsilon}{\epsilon_{t-1}}, \frac{\aleph_t(\epsilon)}{\aleph_t(\epsilon_{t-1})})$ and $(0,1)$. Similar thresholds can be defined when the relative tradeoffs between the need to reduce the threshold and computational expense are weighed differently.
\par
 

We can now state our proposed threshold selection method given the threshold -- acceptance rate curve:
\begin{enumerate}
\item If $t>0$, define $d_{min}$ to be the minimum distance produced by past simulations.
\item Define $\epsilon^*=\argmax_\epsilon \frac{\partial^2 \aleph_t(\epsilon)}{\partial \epsilon^2}$.
\item If $\aleph_t(\epsilon^*) > \delta$ or $\epsilon^*>d_{min}$ (for the case $t>0$), set $\epsilon_t = \epsilon^*$. {\it (Detection of a possible local minimum)} 
\item If $\aleph_t(\epsilon_t)\leq \delta$ and $\epsilon^*\leq d_{min}$, choose $\epsilon_t=\argmin_\epsilon\Delta((\frac{\epsilon}{\epsilon_{t-1}}, \frac{\aleph_t(\epsilon)}{\aleph_t(\epsilon_{t-1})}),(0,1))$.  {\it(Reduction of threshold v.s. Computational expense)}
\end{enumerate}
Of course, the shape of the threshold -- acceptance rate curve is in general unknown but below we show that it is often possible to obtain useful predictions of this curve which allows us to ``guess" near-optimal thresholds. Before we discuss this approach in detail we illustrate its considerable advantages in our toy model.
\afterpage{
\begin{figure}[H]
\center
\includegraphics[width=0.9\textwidth]{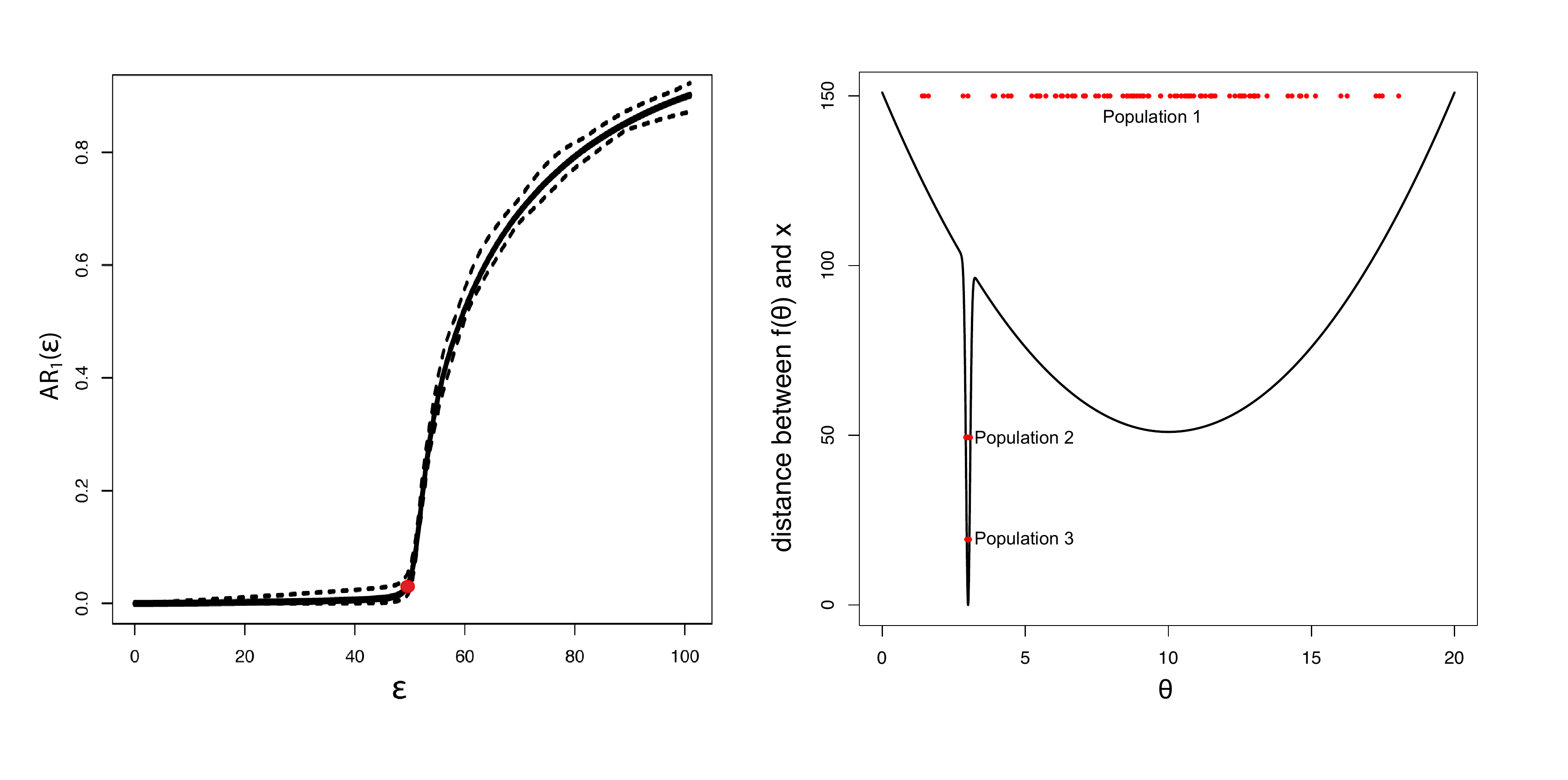}
\caption{\footnotesize{Application of our method to the toy model with a local optimum. (a) The threshold -- acceptance rate curve for the toy model, with solid and dotted lines indicating the mean, maximum and minimum predicted values over 100 runs of our algorithm. The red dot indicates the value $\epsilon^*=\argmax_\epsilon \frac{\partial^2 \aleph_0(\epsilon)}{\partial \epsilon^2}$. (b) Starting from a population spread across both the global and local minimum, a threshold value of approximately $50$ rejects all parameters except those situated in a small interval about the true parameter value of 3. Successive populations refine the distribution about this value.  }}
\label{fig:ToySol}
\end{figure} 
}
\subsection{Toy model redux}
We now revisit the toy model introduced earlier, and illustrated in figure 1. Recall that the existence of a broad local optimum and narrow global optimum, was allowing the ABC SMC inference to converge to a distribution that shared no support with the true posterior. In figure 1b, we examine how likely this is to occur for different fixed quantile schedules, and find that the failure rate increases with the quantile value -- for quantiles of 0.3 and higher, the failure rate is greater than $80\%$. This makes sense, as higher quantile values allow particles from the local optimum region to be accepted for more ABC runs. For each of these runs it is unnecessary to sample from the global optimum in order to reach $N$ accepted particles, and so an opportunity exists to either miss it entirely, or to sample it sparsely enough that it is lost during perturbation. We also find that the number of simulations needed for convergence of the sequence of epsilons, reduces as the quantile increases - that is, the computational expense of failure is lower than success.
\par
The epsilon - acceptance rate curve for this model is shown in Figure \ref{fig:ToySol}a. The shape is sigmoidal with the position of the lower ``elbow'' at $\epsilon \approx 50$, which correctly predicts the minimal distance obtainable from parameters in the local optimum region. Setting epsilon to this value will cause nearly all particles from the local optimum to be rejected, and force particles closer to the true posterior to be sampled. Indeed, application of our method allows ABC SMC to converge to a distribution about the true parameter value every time for this example. One such inference is shown in figure \ref{fig:ToySol}b, where the second population is already restricted to a small interval about the true parameter value. While the fixed quantile methods can be similarly successful e.g. for 0.05, in practice this occurs only with the good fortune of choosing a quantile that selects a threshold appropriately with respect to some unknown critical value (here $50$) sufficiently quickly.

\section{Estimating the threshold -- acceptance rate curve}
The threshold selection scheme described above relies upon knowledge of the threshold -- acceptance rate curve. In this section we suggest a computationally inexpensive way in which it can be approximated.
\par
We first define formally the acceptance rate of the algorithm for round $t>0$ and any threshold, $\epsilon$ by, 
\begin{equation}
\aleph_t(\epsilon) := \int{p_t(x)\mathds{1}\left(\Delta(x,x^*)\le \epsilon\right)}dy
\end{equation}
where $p_t(x)$ is the distribution of the simulated data corresponding to parameters sampled from the last population and perturbed by the kernel $K_t$, i.e.
$$
 p_t(x)=\int q_t(\theta)f(x|\theta)d\theta
 $$
with the perturbed distribution, $q_t(\theta)$, defined via
$$
q_t(\theta):=\sum_{i=1}^N \omega^{(i,t-1)} K_t(\theta|\theta^{(i,t-1)})\;.
$$
A simple way to estimate $p_t(x)$ would be via Monte Carlo approximation, i.e. simulating data from a large sample drawn from $q_t(\theta)$. However, the expense of such a naive approach is generally prohibitive. Here we  use the so-called unscented transform to approximate the distribution $p_t(x)$ of the model output given the distribution $q_t(\theta)$. The unscented transform (UT) \cite{Julier:2000cb}, tells us how the moments of a random variable, $\theta$, are transformed by a non-linear function, $g$. Its computational efficiency and flexibility to the form of the non-linear function, has led to its extensive use in filtering \cite{Wan:2000vj} and smoothing \cite{Briers:2004ug} algorithms, and its increasing popularity as a tool for parameter inference \cite{Quach:2007vv, Liu:2012ej} and uncertainty propagation \cite{Giza:2009ug}. From here we will model our data as a non-linear transformation, $g$, of the parameter $\theta$ with an additive zero-mean noise term. However, the method can be extended to stochastic state space models, with some limitations on the form of the observation model, or without these limitations in an ABC filtering framework.
\par
\begin{figure}
\center
\includegraphics[width=0.6\textwidth]{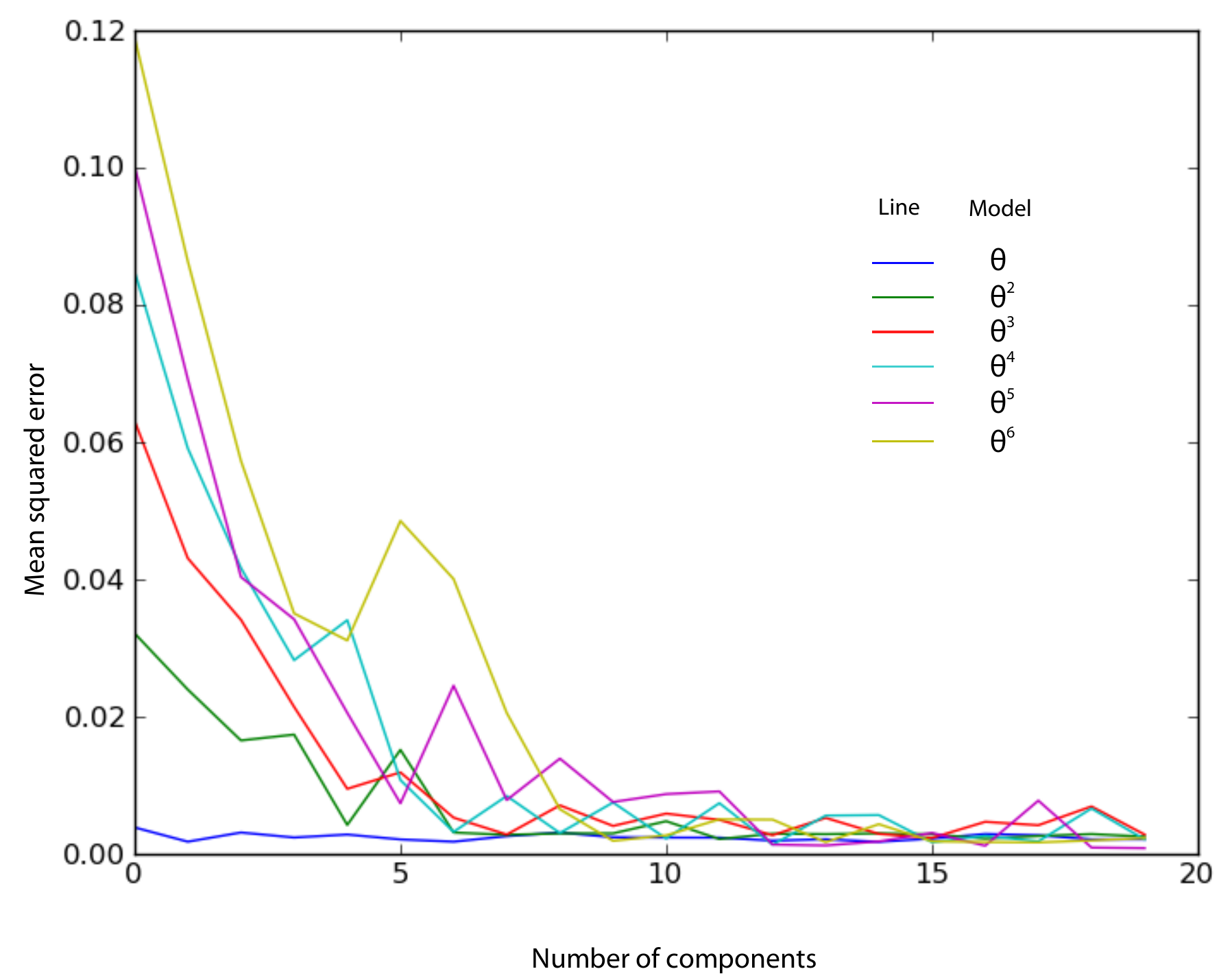}
\caption{\footnotesize{Prediction errors. Plot showing the mean squared error in predicting acceptance rates over 10 values of $\epsilon$ for different numbers of mixture components. Results for models of varying degrees of non-linearity are shown in different colours. }}
\label{fig:errors}
\end{figure}

The unscented transform requires that the perturbed distribution, $q_t(\theta)$, is decomposed into a mixture of Gaussians,
$$
q_t(\theta) \approx \sum_i\alpha_i p_i(\theta)
$$
with each $p_i$ being a Gaussian density that can be fit by an EM algorithm. In general, the greater the number of components the more accurate the acceptance rate approximations become --- we are not only trying to fit the input distributions but allow enough flexibility to approximate a possibly complex, multi-modal output. Indeed we find that increasing degrees of non-linearity in $g$ requires more Gaussian components in order to keep the accuracy at the same level (see figure \ref{fig:errors}).
\par
For each component of the mixture, we use the UT to approximate,
$$
\int f(x|\theta)p_i(\theta)d\theta,
$$
as follows. The first step in the UT algorithm is to determine a set of weighted particles (called sigma-points) with the same sample moments up to a desired order as the distribution $p_i(\theta)$. Here we use a scaled sigma-point set $\{\chi_k\}_{k=0,\ldots 2L}$ that captures both means and covariances \cite{Julier:2002un},
$$
\begin{tabular}{l l l l}
$\chi_0= \mu_\theta$ &  \\
$\chi_k= \mu_\theta + \left[{ \sqrt{(L+\lambda){\Sigma_\theta}}}\right]_{k}$  & $k=1,...,L$ \\
$\chi_k = \mu_\theta - \left[{ \sqrt{(L+\lambda){\Sigma_\theta}}}\right]_{k}$ &  $ k=L+1,...,2L$  \\
\end{tabular}
$$
\noindent where $L$ is the dimension of $\theta$, $\mu_\theta$ and $\Sigma_\theta$ are the mean and covariance of $\theta \sim p_i(\cdot)$, $[A]_{k}$ represents the $k$th column of a matrix $A$, and 
$$
\lambda = \alpha^2(L+\kappa)-L.
$$
The sigma-point weights $\{\upsilon^c_k, \upsilon^m_k\}_{k=0,\ldots 2L}$ are given by,
$$
\begin{tabular}{l l l l}
$\upsilon^m_0 = \frac{\lambda}{L+\lambda}$ &  \\
$\upsilon^c_0 = \frac{\lambda}{L+\lambda}+(1-\alpha^2+\beta)$ &\\
$\upsilon^m_k = \upsilon^c_k = \frac{1}{2(L+\lambda)}$ & $k=1,...,2L$.\\
\end{tabular}
$$
and finally, the parameters $\kappa$, $\alpha$ and $\beta$ may be chosen to control the positive definiteness of covariance matrices, spread of the sigma-points, and error in the kurtosis respectively. While sigma-point selection schemes exist for higher moments \cite{Tenne:2003hj, Julier:1998vm}, they come with significantly increased computational cost.
\par
Once determined, each sigma-point is propagated individually through the function, $g$, and the mean and covariance of the transformed variable, $g(\theta)$, can be estimated using the update equations,
\begin{eqnarray}
\mu_{g(\theta)} &\approx & \sum^{2L}_{k=0}\upsilon^m_kg(\chi_k)\\
\Sigma_{g(\theta)} &\approx & \sum^{2L}_{k=0} \upsilon^c_k(g(\chi_k)-\mu_{g(\theta)})(g(\chi_k)-\mu_{g(\theta)})^T .
\end{eqnarray}
We denote the resulting approximate probability density function for $g(\theta)$, for mixture component $i$ by $\mathcal{U}_{p_i}(x)$.
\par
By matching terms in the Taylor expansions of the estimated and true values of these moments, it can be shown that the above algorithm is accurate to second order in the expansion. More generally, if the sigma-point set approximates the moments of $\theta$ up to the $n^{th}$ order then the estimates of the mean and covariance of $g(\theta)$ will be accurate up to the $n^{th}$ term \cite{Julier:2000cb}. Crucially, the number of points required ($2L+1$ for this scheme) is much smaller than the number required to reach convergence with Monte-Carlo methods.
\par
Given the $\mathcal{U}_{p_i}(x)$, for each mixture component, we can approximate the distribution of the output $x$ as follows,
\begin{eqnarray}
p_t(x) &\approx& \sum_i\alpha_i\int f(x|\theta)p_i(\theta)d\theta\\
        &\approx& \sum_i\alpha_i\; \mathcal{U}_{p_i}(x) \label{UTmixtureapprox}.        
\end{eqnarray}
Samples $\{x_j\}_{ j=0,...,M}$ from the mixture of Gaussians distribution in equation~\eqref{UTmixtureapprox} may then be used as an inexpensive proxy for ABC simulations, and the acceptance rates can be estimated as,
\begin{equation}
\aleph_t(\epsilon) \approx \frac{1}{M}\sum^M_{j=1} H_{\epsilon}(\Delta(x_j,x^*))\; \label{ARapprox}
\end{equation}
where,
\begin{equation}
H_{\epsilon}(\Delta(x_j,x^*)) = \frac{1}{\left(1+e^{-k\left(\frac{\Delta(y,y^*)}{\epsilon}-1\right)}\right)}
\end{equation}
is used as a smooth approximation to the ``accept and reject'' indicator function, with $k$ controlling the severity of the step. The smooth approximation is necessary for estimating the critical value,
$$
 \frac{\partial^2 \aleph_t(\epsilon)}{\partial \epsilon^2} =  \frac{1}{M}\sum^M_{j=1} \frac{\partial^2 H(\Delta(x_j,x^*))}{\partial \epsilon^2}
$$
of the proposed threshold selection scheme.
\par
In summary, the ABC SMC acceptance rate may be approximated for any threshold value at the beginning of each round $t>0$ of the algorithm using the steps:
\begin{enumerate}
\item generate a population of perturbed particles, sampling from $\{\theta^{(i,t-1)},\omega^{(i,t-1)}\}_{1\leq i\leq N}$ and perturbing each particle independently with $K_t$,
\item fit a Gaussian mixture model to the perturbed population,
\item estimate $p_t(x)$ using the unscented transform independently for each component $p_i$ of the Gaussian mixture,
\item estimate acceptance rates for different threshold values according to equation~\eqref{ARapprox} sampling from $\sum_i\alpha_i\; \mathcal{U}_{p_i}(x) $.
\end{enumerate}

\section{Applications to biological models}
We now contrast our adaptive method to various fixed quantile threshold schedules in the context of two biological dynamical systems. We consider two criteria: firstly, the total number of simulations required to reach a pre-chosen threshold value; and secondly, the proportion of repeat runs that fail, i.e. get stuck in a local minimum, or fail to reach the threshold in a given (very long) period of time.

\subsection{Computational expense: Quantiles vs UT}
The repressilator has become a classic example of a synthetic biological oscillator \cite{Elowitz:2000gg}. It consists of six species (three mRNAs, $(m_i)$, and their protein products, $(p_i)$),  with regulatory links between them forming a single feedback loop --- each protein inhibits the production of the next protein's mRNA. The dynamics of the species concentrations are governed by the first order differential equations,

\begin{eqnarray}
\frac{dm_1}{dt} &=& -m_1 + \frac{\alpha}{1+p^n_3} + \alpha_0,\\
\frac{dp_1}{dt} &=& -\beta(p_1 - m_1),\\
\frac{dm_2}{dt} &=& -m_2 + \frac{\alpha}{1+p^n_1} + \alpha_0,\\
\frac{dp_2}{dt} &=& -\beta(p_2 - m_2),\\
\frac{dm_3}{dt} &=& -m_3 + \frac{\alpha}{1+p^n_2} + \alpha_0,\\
\frac{dp_3}{dt} &=& -\beta(p_2 - m_2),
\end{eqnarray}
\noindent where $\theta = ( n, \beta, \alpha,\alpha_0)$ is the parameter vector to be inferred. We set the initial species concentrations to $(m_1,p_1,m_2,p_2,m_3,p_3)=(0.0,2.0,0.0,1.0,3.0)$, and generate some data by simulating the model with $\theta = (2.0,4.0,1000.0,1.0)$, and ``observing'' the state of $p_1$ at time-points $(4.0,8.0,...,20.0)$, subject to some small added zero-mean Gaussian noise with covariance $0.01I$. With Gaussian prior distributions that encompass the true parameter values, we perform ABC SMC, choosing thresholds according to our method and a range of fixed quantile threshold schedules. The inferences are repeated $10$ times for each method, and stopped once a round of ABC has been completed with a threshold value below a pre-chosen challenging threshold - in this case, $35$.
\par
\begin{figure}[H]
\center
\includegraphics[width=0.7\textwidth]{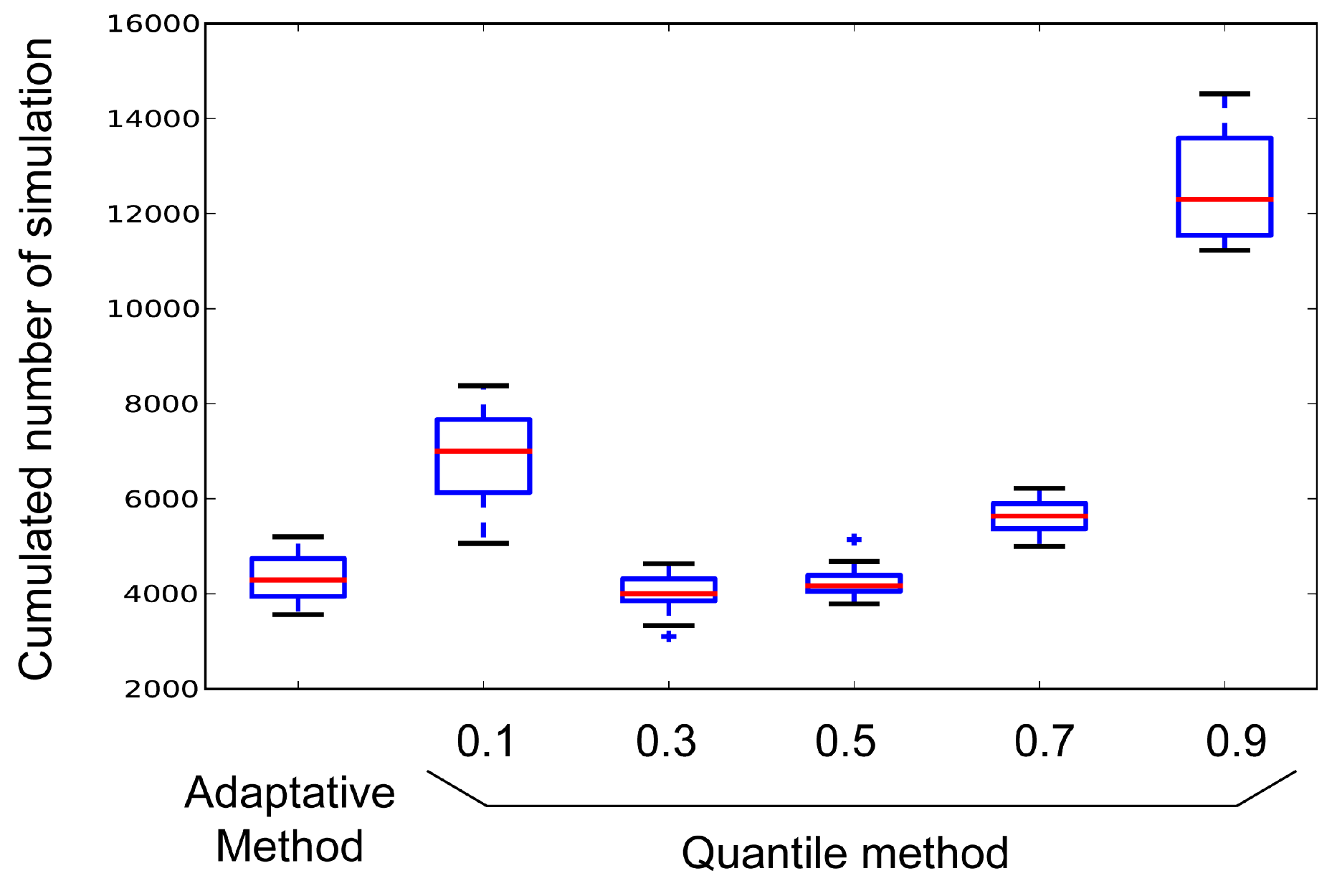}
\caption{\footnotesize{Box plots showing the total number of simulations required to reach a threshold value of 35 (with the L2 distance function), for the Repressilator model and different threshold schedules. An order of magnitude difference exists in the computational expense of the best and worst performing quantile methods. Our adaptive method performs comparably to the best fixed quantile schedules.  }}
\label{fig:Repress}
\end{figure}

A comparison of the performance of each method is shown in figure \ref{fig:Repress}. The computational expense of the quantile methods is found to vary significantly and in non-linear fashion, with the best performing quantile ($0.3$) over three times cheaper than the worst ($0.9$). This highlights the difficulty of choosing fixed threshold values that perform well. Our method scores similarly to the best fixed schedules (at approximately $4000$ simulations), which suggests that it is successfully reducing the computational cost for this inference. 

\subsection{Oscillations and local minima}
\begin{figure}[H]
\center
\includegraphics[width=0.8\textwidth]{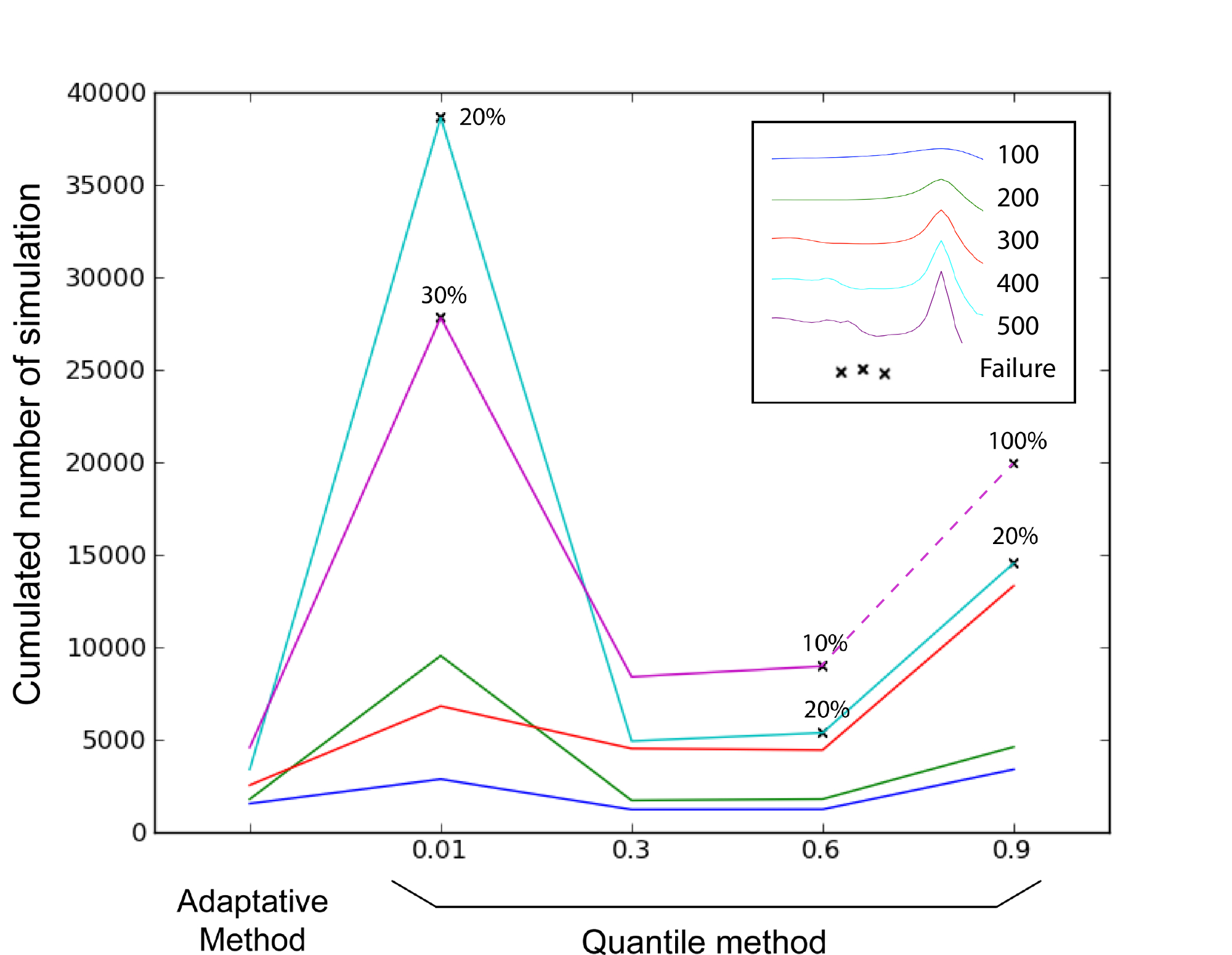}
\caption{\footnotesize{Plots showing the performance of our adaptive method and different fixed quantile schedules for parameter inference on the Hopf bifurcating system. For each data set and method, 10 inferences were performed. As shown in the legend, colours indicate the number of data points used for the inference, and crosses mark failures of some inferences to reach a fixed threshold value (scaled according to the number of points used), within 100000 simulations; the percentage failure is also shown. The legend also shows the shapes of the likelihood surface for each dataset, with the peak around the true parameter value narrowing as the set size increases. Note that the computational expense varies least across datasets for our method, which also suffers no failures. In comparison, the 0.9 quantile schedule always fails when using 500 data points.}}
\label{fig:Hopf}
\end{figure}

The model below represents the simplest biochemical reaction system that permits a Hopf bifurcation  \cite{Kirk:2008ws,Wilhelm:1995uc}. It can be shown that this system, described by,
\begin{align*}
\frac{dx}{dt} &=(Ak_1-k_4)x-k_2xy\\
\frac{dy}{dt}&= -k_3y+k_5z\\
\frac{dz}{dt} &= k_4x-k_5z \; ,
\end{align*}
\noindent where, $x$, $y$, $z$, represent the concentrations of three reactants, $k_i$, are the reaction rates, and, $A$, is the fixed concentration of a fourth reactant, displays a limit cycle for $Ak_1=k_3+k_4+k_5$. Further, when the true value of $Ak_1$ is greater than the critical value, $k_3+k_4+k_5$, the bifurcation has an effect on the likelihood of producing a global maximum and broader local maximum, with the regions becoming more defined for larger data sets \cite{Kirk:2008ws}. This is illustrated in the legend of figure \ref{fig:Hopf}, where the shapes of the log-likelihood (with respect to $Ak_1$) are shown for data sets of size $T = (100,200,...,500)$, with fixed $k_i = 1$ and the true value of $Ak_1$ set at 5.5. Values of $Ak_1$ below the bifurcation point ($Ak_1=3$) are seen to have log-likelihood values which, in the case of larger $T$, are greater than those for values of $Ak_1$ above the bifurcation point that do not belong to a small interval about the true value. 
\par
We repeat ABC SMC inferences 10 times for each of the data sets and for each threshold choice scheme. The total number of simulations needed to reach a target threshold value of $\sqrt{80T}$ (scaled according to the size of the data set), is recorded unless this grows above 100,000, in which case the inference is considered to have failed. By varying $N$ we are able to examine the adaptability of our threshold choice method to different likelihood shapes and, moreover, perform a ``stress test'' by controlling how challenging it is to avoid the local maximum. 
\par
Results comparing our strategy to various fixed quantile schedules are shown in figure \ref{fig:Hopf}. We find that in all cases, the expense of our method is comparable or cheaper than the best performing fixed quantile schedule, and further that the variability in cost between different data sets is smallest for our method. Moreover, our method is successful in all cases, while the fixed quantile schedules (with the exception of $0.3$) suffer failures for $T \geq 300$; in one case  (for $\alpha=0.9$) this happens for every repetition. For the larger quantiles, these failures are caused by the accepted population becoming trapped in the interval $(0,3)$, while for the $0.01$ quantile, the reduction in threshold can be too severe, which leads to a very low acceptance rate and computationally overly expensive inference. These observations suggest that our strategy is successfully adapting to the different likelihood shapes; thresholds are chosen that balance the need to avoid the local optima with minimising computational expense.

\section{Discussion}
Knowing the relationship between the acceptance rate and the $\epsilon$ threshold schedule applied in ABC SMC has obvious implications for the efficiency and computational affordability of ABC inference in complex inference tasks; obviously, ABC schemes should really only be used in cases where conventional likelihood-based inferences fail. Clearly, however, this relationship is unknown but as we have shown naive reliance on threshold schedules determined adaptively from quantiles of previous populations are fraught with a number of problems. The most commonly encountered problem is that gentle reductions in thresholds, $\epsilon_t$, between populations have a tendency to lead populations of particles to diffuse into regions of low posterior probability. This, we feel is a powerful argument against quantile-based adaptive schemes, in particular those were the computational cost is fixed, i.e. where the number of simulations is specified and a fixed fraction of the simulated particles with the smallest distances are used to make up the intermediate population. The attraction of fixed (or controlled) computational burden comes with the high risk of convergence to spurious and biased ``posterior" distributions. 
\par
Here we have shown how the UT can be employed to predict the shape of the threshold -- acceptance rate curve, and select a (near-)optimal threshold value. This approach is superior to the selection of a fixed quantile criterion for the threshold choice, as the optimal quantile depends critically on the problem at hand. The UT is an ancillary or supporting statistical inference step which allows us to fine-tune the technical parameters of the inference process; in no way does this interfere with conventional Bayesian practice or conventions. It merely allows us to overcome some of the limitations inherent to the ABC approach (where indicator functions replace continuous probability measures).
\par
We have here focussed on dynamical systems where the observed data are compared directly to the simulations, rather than summary statistics of real and observed data. Extension of this approach to inference using appropriate summaries \cite{Joyce:2008p20301,Nunes:2010dv,Barnes:2012ux} is in principle straightforward as the UT only aims to predict the shape of the threshold -- acceptance rate curve. The remaining parts of the algorithm are not affected by this in principle, although in practice the Gaussian mixture model and other factors affecting efficiency and accuracy of the predictions may need to be considered carefully. 
\par
It has to be kept in mind that at the moment we adopt a greedy procedure and predict only the next threshold. Providing a global choice of the $\epsilon_t$ threshold for all $t$ is a much harder inference task. Although the overall number of simulations in the ABC SMC scheme (once all $\epsilon_t$ are determined) may be less than the number of simulations required by our greedy approach, we believe that the computational burden and complications inherent in determining global  schedules are prohibitive. 

\bibliography{bib}
\end{document}